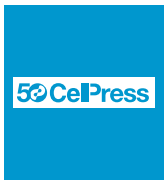
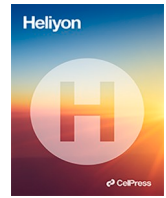

Research article

# Hemodynamical behavior analysis of anemic, diabetic, and healthy blood flow in the carotid artery

Hashnayne Ahmed [a,b,*], Chinmayee Podder [a]

[a] *Department of Mathematics, Faculty of Science & Engineering, University of Barishal, Barishal, 8200, Bangladesh*
[b] *Department of Mechanical and Aerospace Engineering, University of Florida, Gainesville, FL, 32611, USA*

ABSTRACT

The influence of blood rheology on hemodynamic parameters is investigated using Computational Fluid Dynamics on blood flow through the human carotid artery. We performed three-dimensional modeling and simulation to study blood flow through the carotid artery, which is divided into internal and exterior parts with a decreased radius. The blood flow was classified as basic pulsatile to simulate the human heart's rhythmic pulses. For hemodynamic modeling viscosity of the fluid, the Carreau model was utilized with four distinct blood instances: Anemic, diabetic, and two healthy blood types. The boundary conditions with Carreau viscosity were applied using the Ansys Fluent simulator, and the governing equations were solved using the finite volume technique. Different time steps were tested for their impact on wall deformation, strain rate, blood velocity, pressure, wall shear, and skin friction coefficient. The hemodynamical parameters were calculated using many cross-sectional planes along the artery. Finally, the impact of the four types of blood cases listed above was investigated, and we discovered that each blood case has a substantial impact on blood velocity, pressure, wall shear, and strain rate along the artery.

## 1. Introduction

Blood plays a key role in regulating the body's internal systems and functioning the living organism to adjust its internal environment into a stable state position. It performs many functions within the body, such as providing oxygen and nutrients to tissues, removing waste materials, transporting of hormones and other signals, regulating the body's immune and hydraulic functions, temperature and pH, endocrine and digestion systems throughout the whole body [41]. So, major health conditions that affect the blood flow can be life-threatening to humankind. Atherosclerosis is one of the most common cardiovascular diseases blocking blood's regular movement, which causes increased morbidity. Even though atherosclerosis is a complex disease, it predominantly attacks artery bifurcation walls [1]. Its concentrated nature is caused by local hemodynamic characteristics such as wall shear stress and flow recirculation in specific locations [2]. Atherosclerosis of the large arteries is one of the leading causes of ischemic stroke, as well as a major cause of mortality and disability. Carotid stenosis and blockage are intimately linked to weakened brain hemodynamics. Transient ischemic attack is one of the symptoms [3]. The degree of internal carotid artery stenosis does not affect the early risk of stroke. Stroke can also occur because of plaque rupture in a stenosed carotid artery, which is a favorable site for the development of atherosclerosis due to arterial bifurcations that disrupt and complicate flow [4]. Because of the relevance of hemodynamic parameters such as blood pressure and wall shear stress in the development and progression of cardiovascular disease, a full understanding of carotid artery hemodynamics is required [3].

Blood is a non-Newtonian fluid that contains platelets, white blood cells, red blood cells, and other components [5], however, researchers have considered blood as a Newtonian fluid because in large vessels (diameter varying from 1 to 3 *mm*) the shear rates are higher than 100 $s^{-1}$ [6–9]. In cardiovascular biomechanics, the structural modeling of the artery is thoroughly studied [10,11]. Assume that the arterial material is linearly elastic. This is a straightforward technique to represent an artery. This assumption has the benefit of being computationally simple, and the findings generated by this method are consistent with experimental data within the physiological pressure range [12]. This approach is beneficial in patient-specific fluid-solid interaction simulations and is employed in many situations when the artery is linearly elastic in the physiological pressure range [13]. The approach allows for the simulation of the wall's stiffening behavior and incompressibility under high-strain circumstances. More advanced arterial wall models are also used, considering the viscoelastic, inhomogeneous features of the arterial wall due to its compositions [14]. Because the constituents of sick arteries are more diverse and intricate, such models are desirable [15]. However, because vascular information such as composition, fiber orientation, lipid accumulation locations, and calcification and its characteristics is difficult to collect, the use of these material models in patient-specific investigations is limited. Many researchers have utilized various elasticity models in their studies and found that when vascular tissue is characterized by a nonlinear model, it behaves more rigidly than when it is described by a linear elastic model [16] and the adjustments aren't significant. As a result, the artery wall is treated as a linear elastic material in patient-specific fluid-solid interaction simulations. When compared to a more realistic flexible wall method, the rigid wall assumption overestimates the wall shear stress by up to 50%, resulting in certain qualitative and quantitative deviations. As a result, elastic artery models are more appropriate [17].

Hypertension is one of the risk factors of cardiovascular diseases such as stroke. The effect of hypertension on aneurysms and stenosed arteries has been widely researched. According to the researchers, hypertension increases wall shear stress and deformation in aneurysm regions, which is produced by high stress-strain situations. They also observed that hypertension increases stenosis severity in the carotid artery [18,19]. This increases the angular phase difference between wall shear stress and circumferential stress waves at the stenosis throat, as well as the negative wall shear stress and oscillatory shear in the post-stenosis region [35,36,39].

Meanwhile, the carotid artery carries oxygenated blood to the brain and face. The common carotid artery is divided into two arteries, one with a bigger diameter known as the internal carotid artery and one with a smaller diameter known as the external carotid artery. Any blockage in the carotid artery can cause ischemia in the brain, leading to a stroke. As stroke is the second leading cause of death worldwide, it is vital to monitor blood flow in the carotid artery [20]. Ischemic and hemorrhagic strokes are the two most common forms of strokes. According to statistics, 87% of strokes are ischemic, 10% are intracerebral hemorrhage, and 3% are subarachnoid hemorrhage [21].

Diabetes causes an increase in blood viscosity, which causes greater resistance to blood flow, hemoglobin or oxygen dissociation, capillary fragility, platelet function, and vascular occlusion, which can lead to tortuosity and multiple stenoses [22–25]. Anemia (low red blood cell count) leads blood to behave like a Newtonian fluid. Low viscosity blood has less wall shear stress, which results in lower pressure. It is, however, problematic for the heart since it must work harder to compensate for the low red blood cell count to deliver an adequate amount of oxygenated blood [26].

There are several techniques for stimulating blood flow in arteries; however, computer simulations provide a good, cost-effective, and non-invasive way of studying flow within vessels [34]. These simulations can display practically any parameter that medical practitioners could find relevant. This study looks at the effects of diabetes and anemia on carotid artery diseases, namely ischemic stroke. Consequently, the 3*D* model of the carotid artery was utilized in the Ansys FLUENT to examine the impact of stenosis on many metrics for diabetic, anemic, and two healthy types of blood. In a stenosed carotid artery, Attia and Eldosoky looked at three types of blood behaviors for blood velocity and pressure [33]. They discovered that when there is stenosis, the blood flow velocity rises, and it increases much more in diabetic blood and that the high oscillation of velocity patterns implies a higher chance of stenosis formation in certain areas.

To understand the impact of the blood types in the artery and arterial wall, numerous hemodynamic behaviors have been investigated in this article.

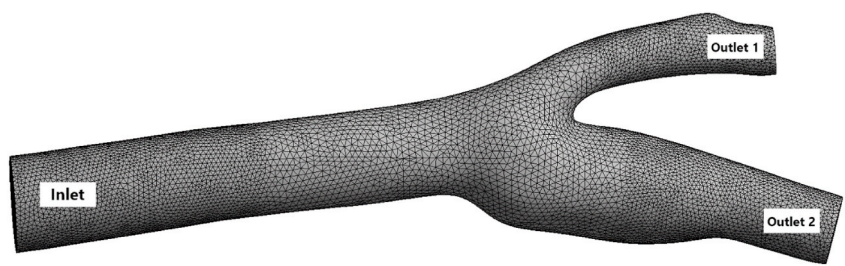

**Fig. 1.** Carotid artery with the inlet (common carotid artery), outlet 1 (external carotid artery), and outlet 2 (internal carotid artery).

## 2. Physical model

A 3D model of a carotid artery with its branches is taken for this study as shown in Figs. 1 and 2. The carotid artery has three parts that show the artery in Y-shape [30]. The common carotid artery (CCA) is the inlet, which is divided into two arteries: the external carotid artery (ECA) and the internal carotid artery (ICA). Because the size of the carotid artery and its tributaries vary from person to person, we base our assumptions on previously utilized data [31,32]. The geometry is picked at random and subsequent adjustments are made with SolidWorks.

## 3. Mathematical formulation

The carotid artery blood flow is pulsatile, cyclic, non-Newtonian, and incompressible in this article. Because blood is a non-Newtonian fluid, its viscosity coefficient is not constant [27]. The fluid mass and momentum conservation can be described by the following Eq. (1) and Eq. (2) [28,29]:

$$\nabla.\pmb{v} = 0 \tag{1}$$

$$\rho\left(\frac{\partial \mathbf{v}}{\partial t} + \mathbf{v}(\nabla.\pmb{v})\right) = -\nabla p + \mu \nabla^2 \pmb{v} + f \tag{2}$$

$$\mu_{eff}(\gamma) = \mu_{inf} + (\mu_0 - \mu_{inf})\left(1 + (\lambda\gamma)^2\right)^{\frac{n-1}{2}} \tag{3}$$

We used the Carreau blood viscosity model [28,29] described by Eq. (3) to explore different blood cases including healthy, diabetic, and anemic blood samples with the properties described in Table 1. By altering the model's viscosity parameters to match various hematocrit levels in the blood that reflects the circumstances for the considered cases. Blood with a hematocrit level of 45 percent is considered healthy blood in the following table; however, a diabetic and anemic patient's blood hematocrit count is 65 percent and 25 percent, respectively [22]. This study considers different cases and acknowledges that hypotheses may vary with age, gender, and patient-specific factors.

The arterial wall shear stress may be estimated using the following Eq. (4) [37]:

$$\tau_{wss} = -\mu \left(\frac{\partial v}{\partial r}\right)_{wall} \tag{4}$$

And the skin friction co-efficient is defined using the following Eq. (5) [38]:

$$C_f = \frac{\tau_{wss}}{0.5\ \rho v^2} \tag{5}$$

For wall boundary conditions, artery wall characteristics are considered. Sinnottet et al. suggested the following model (Eq. (6)) for pulsatile and cyclic blood flow [24]:

$$v_{inlet}(t) = \begin{cases} 0.5\ \sin[4\pi(t + 0.0160236)], 0.5n < t \leq 0.5n + 0.218 \\ 0.1, 0.5n + 0.218 < t \leq 0.5(n+1) \end{cases} \tag{6}$$

where $n = 0, 1, 2, 3, 4, \ldots\ldots\ldots$.

Throughout each interval, the pulsatile profile is a combination of two phases. The velocity at the intake fluctuates sinusoidally throughout the systolic period. The sine wave has a peak velocity of $0.5\ m/s$ and a minimum velocity of $0.1\ m/s$ during the systolic period. And each cycle lasts 0.5 seconds if the heartbeat rate is 120 beats per minute. A healthy person's systolic pressure is about

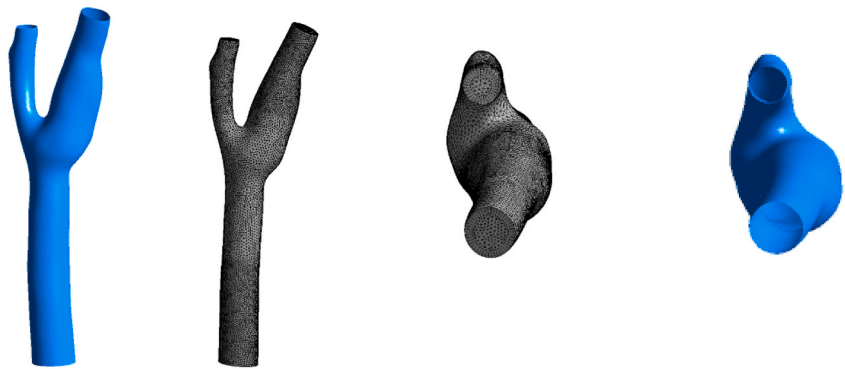

**Fig. 2.** Schematic view of the considered carotid artery.
Grabcad link of the geometry: https://grabcad.com/library/carotid-bifurcation.

**Table 1**
Experimental data for hematocrit counts and viscosity.

| Properties | Anemic Blood [33] | Diabetic Blood [33] | Healthy Blood (Case 1) [33] | Healthy Blood (Case 2) [34] |
| --- | --- | --- | --- | --- |
| Power-law index, $n$ | 0.33 | 0.39 | 0.48 | 0.3568 |
| Constant time, $\lambda$ $(s)$ | 12.448 | 103.09 | 39.418 | 3.313 |
| $\mu_{inf}$ $(Pa.s)$ | 0.00257 | 0.00802 | 0.00345 | 0.0035 |
| $\mu_0$ $(Pa.s)$ | 0.0178 | 0.8592 | 0.0161 | 0.056 |
| Hematocrit count | 25% | 65% | About 45% | |

120 $mmHg$, while their diastolic pressure is around 80 $mmHg$. As a result, we pick 100 $mmHg$ (about 13332 $Pa$) as the static gauge pressure at the outlets, based on the average pressure of the two phases.

## 4. Grid independence and validation

Based on the examination of multiple grid conditions, the grid independence test is a procedure used to discover the ideal grid condition with the fewest number of grids without causing a difference in numerical results. That is, the purpose of this test is to determine the best mesh size for the simulation. The product of strain and cell wall distances is used to compute wall deformation. Fig. 3 depicts the variance in average wall deformation for various grid sizes.

The rate of blood flow, or velocity, is inversely proportional to the total cross-sectional area of the blood arteries. The velocity of flow reduces as the total cross-sectional area of the tubes rises. The capillaries have the slowest blood flow, which allows for the exchange of gases and nutrients. As previously indicated, blood types and artery types have a significant impact on average velocity. The fluctuation in average velocity at outlet 1 (internal) and outlet 2 (external) for different grid sizes is shown in Fig. 4.

The force per unit area imposed by a solid barrier on a fluid in motion (and vice versa) in a direction on the local tangent plane is known as wall shear stress. The fluctuation of average and maximum wall shear stress for different grid sizes is shown in Fig. 5.

Considering the above studies, an optimum mesh size of 514832 elements for the selected geometry is chosen for this study with a minimum triangular element size of 0.3 $mm$.

To choose an optimum time step for the fluid-solid interaction-based simulation, a transient time-step sensitivity test is performed. Transient analysis is carried out for different time steps starting with 25, 50, 100, and 200 steps with a step size of 0.032s, 0.016s, 0.008s, and 0.0064s respectively, and average velocity and average wall shear stress are plotted at different time steps as shown in Fig. 6. Based on the time-step sensitivity test, fluid-solid interaction-based simulations with 50-time steps with a step size of 0.008 $s$ are performed. This selection strikes a balance between capturing important flow dynamics and minimizing numerical errors, leading to relatively lower velocity and wall shear stress values compared to lower time steps.

## 5. Results and discussions

Computational fluid dynamics has been used by biomedical researchers to simulate and understand the physical factors that lead to the creation and evolution of hemodynamic diseases [35]. The passage of blood through the arteries is referred to as blood flow, and blood flow is inversely proportional to the artery's area. The velocity of blood flow decreases as the cross-sectional area of the artery grows, and it increases if plaque is deposited in the artery, causing the cross-sectional area of the artery to drop [36]. The force that blood produces on the arterial wall as it pumps blood through the artery is known as pressure on the arterial wall.

All the simulations were performed with Ansys FLUENT 19.4, while the designs were modified with the SolidWorks program. A user-defined function is utilized for time-dependent modeling needs.

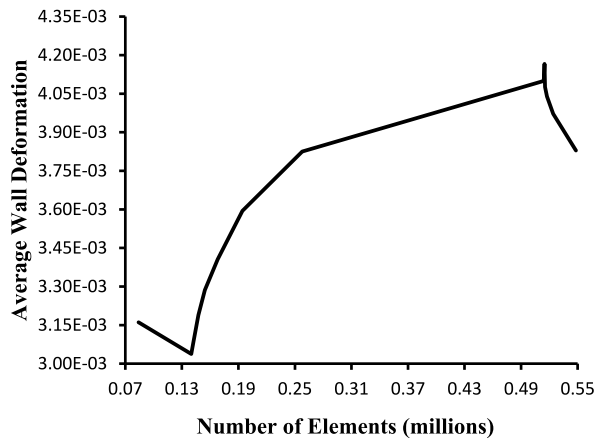

**Fig. 3.** Grid independence based on wall deformation.

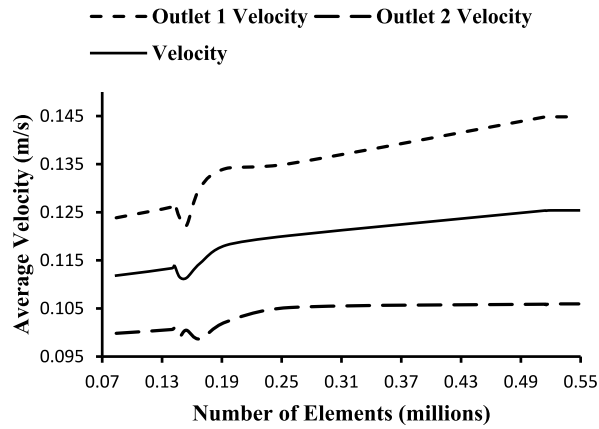

**Fig. 4.** Grid independence based on average outlet velocity.

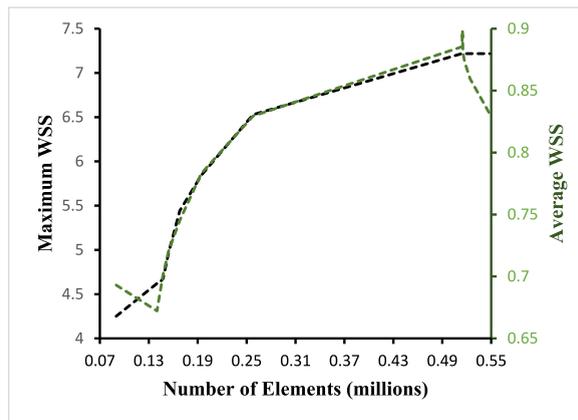

**Fig. 5.** Grid independence based on wall shear stress.

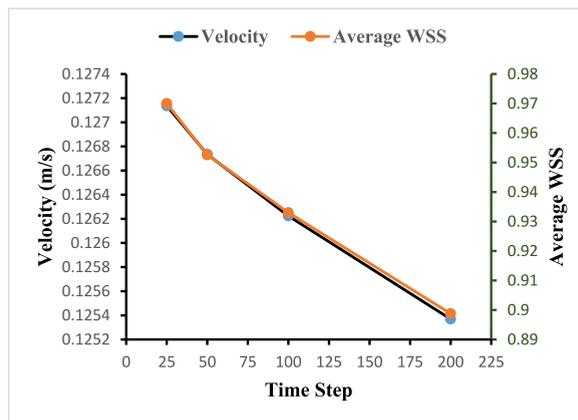

**Fig. 6.** Time step sensitivity test for the optimal time step.

The influence of flow dynamics in the carotid artery is studied using several flow parameters such as wall deformation, strain rate, flow velocity, pressure, wall shear, and skin friction coefficient for various time steps, cross-sectional planes along the artery, and blood types.

## 5.1. Effects of different time steps on the hemodynamic behavior

In this section, Carreau blood flow is considered, and hemodynamic parameters along the carotid artery such as wall deformation, wall shear stress, and flow velocity are explained at different time intervals of the cardiac cycle, and different blood pressure conditions.

The wall deformation of the carotid artery is depicted in Fig. 7 at various time steps. The deformation increases in lockstep with the increase in blood pressure during peak systole, as expected. The artery expands because of the increased distortion, reducing wall shear stress and increasing the recirculation zone. Due to the stored elastic energy in the elastic artery wall, a huge recirculation zone develops at the non-divider wall when flow decelerates after the peak.

From Fig. 7, we can see that the average wall deformation is increasing with the increased time steps, and the average wall deformation is linearly increasing. The highest value of the average wall deformation is 0.034576959 obtained at $t = 0.4s$ and the lowest value is 0.006222495 obtained at $t = 0.08s$. The minimum wall deformation line shows frequently changing characteristics indicated in violet color with circular marks in Fig. 7. The highest value of the minimum wall deformation is $5.64304 \times 10^{-5}$ obtained at $t = 0.2s$ and the lowest wall deformation is $1.19 \times 10^{-5}$ obtained at $t = 0.12s$. The wall deformation of this model is calculated using the average cell wall distance of 0.0011427, multiplication by the deformation value for different time steps.

The change in strain or deformation of a material over time is referred to as strain rate. The following Fig. 8 shows the variations of the average and minimum strain rate concerning the time steps. The average strain rate is continuously increasing with the time steps of $t = 0.08s$ to $t = 0.4s$. The highest value of the average strain rate is $75.6475\ s^{-1}$ obtained at $t = 0.4s$ and the lowest value of the average strain rate is $68.0679\ s-1$ obtained at $t = 0.08s$. Again, the highest value of the minimum strain rate is $0.246917 s^{-1}$ obtained at $t = 0.2s$ and the lowest value of the minimum strain rate is $0.0870958\ s^{-1}$ obtained at $t = 0.12s$. The average and minimum strain rate line shows similar variations to the average and minimum wall deformation line as the average cell wall distance is the same for all the time steps. The maximum strain rate for all the time steps is $1522.73\ s^{-1}$ and the maximum deformation is 0.6961. The black line with the triangle marks indicates the average value and the violet line with the circle marks indicates the minimum value of the hemodynamical parameters.

The black line with the triangle marks indicates the average value and the dark red line with the circle marks indicates the maximum value of the hemodynamical parameters for Figs. 9 and 10. The blood flow velocity varies inversely with the total cross-sectional area of the blood vessels. As the total cross-sectional area of the vessels increases, the velocity of flow decreases. Blood flow is quickest in the center of the vessel and slowest toward the vessel's wall. The average and maximum velocity lines in Fig. 9 exhibit comparable fluctuations for different time steps. The highest value of the average velocity is $0.100446\ ms^{-1}$ obtained at $t = 0.28s$ and the lowest value of the average velocity is $0.0989663\ ms^{-1}$ obtained at $t = 0.08s$. Again, the highest value of the maximum velocity is $0.228956\ ms^{-1}$ obtained at $t = 0.4s$ and the lowest value of the maximum velocity is $0.218118\ ms^{-1}$ obtained at $t = 0.08s$. The lowest value of the minimum velocity is $1.06 \times 10 - 5\ ms^{-1}$ obtained for the time step $t = 0.28s$ near the boundary walls.

Most of this pressure results from the heart pumping blood through the circulatory system. Blood pressure is the force exerted by the blood against the unit area of the vessel's wall. Increasing arterial pressure decreasing vascular resistance whereas decreased arterial pressure increases resistance that means elastic vessels gradually collapse due to reduced pressure. If the arterial pressure reduces to a certain below level, the blood vessels completely collapse [42]. Fig. 10 displays the variational change of the inlet pressure of 13332 Pa for different time steps. The highest value of the average pressure is 13360.1 Pa obtained at $t = 0.08s$ and the lowest value of the average pressure is 13358.1 Pa obtained at $t = 0.4s$. Again, the highest value of the maximum pressure is 13393.7 Pa obtained at $t = 0.08s$ and the lowest value of the maximum pressure is 13390.8 Pa obtained at $t = 0.4s$. The value of the minimum pressure is 13331.8 Pa obtained for all the time steps. Fig. 10 shows the decreasing behavior of the pressure with the increased value of the time steps.

In the study of atherosclerosis in arteries, wall shear stress is one of the effective parameters for early diagnosis and will also be introduced in clinical observation. Vessel wall thickening due to the accumulation of lipids and inflammatory cells which is known as plaque formation. This is strongly linked to wall shear stress [40]. Plaque rupture can be caused by high wall shear stress, while plaque

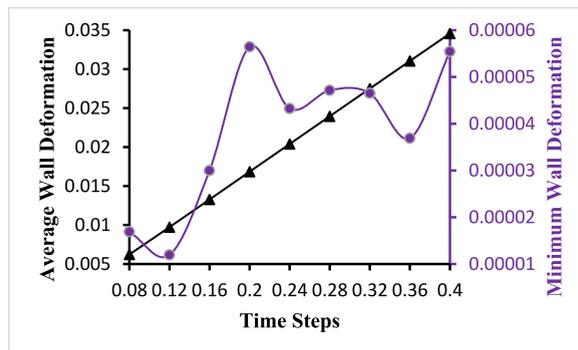

**Fig. 7.** Average and minimum wall deformation variations for different time steps.

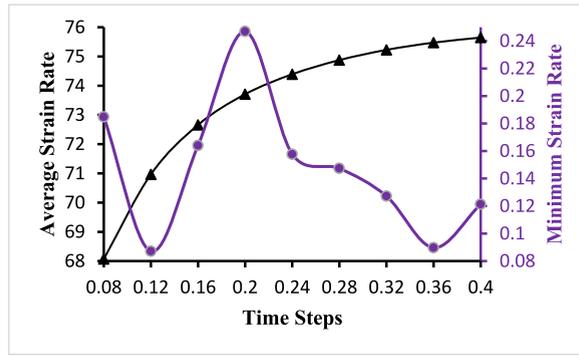

**Fig. 8.** Average and minimum strain rate variations for different time steps.

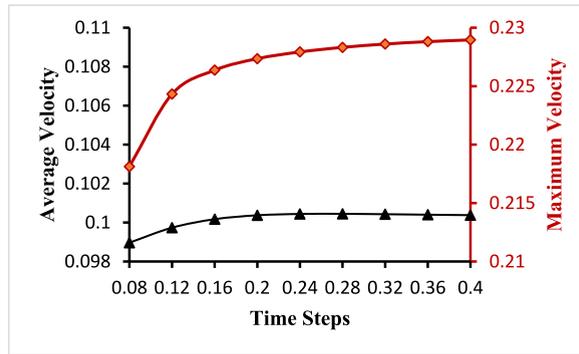

**Fig. 9.** Average and maximum velocity variations for different time steps.

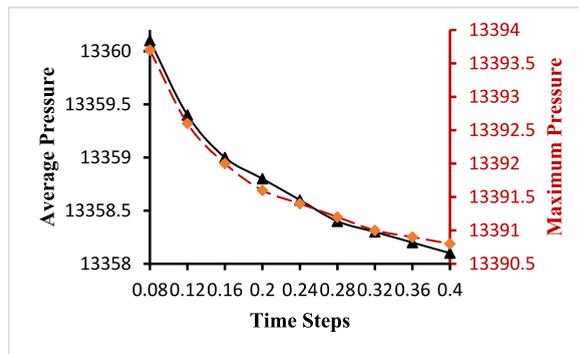

**Fig. 10.** Average and maximum pressure variations for different time steps.

development is caused by low and fluctuating wall shear stress.

Fig. 11 marked with the triangle shows the change of the average wall shear for different time steps in the fluid domain. It shows the continuous decrease of the average wall shear with the time steps. The highest value of the average wall shear is 0.923393 $Pa$ obtained at $t = 0.08s$ and the lowest value of the average wall shear is 0.861389 $Pa$ obtained at $t = 0.4s$. The value of the maximum wall shear is 7.18665 $Pa$ for all the time steps and the value of the minimum wall shear is 0 $Pa$ for all the time steps.

The following Fig. 12 marked with the triangle shows the change of the average skin friction coefficient for different time steps in the fluid domain. It shows the continuous decrease of the average skin friction coefficient with the time steps. The highest value of the average skin friction coefficient is 0.174457 obtained at $t = 0.08s$ and the lowest value of the average skin friction coefficient is 0.162756 obtained at $t = 0.4s$. The value of the maximum skin friction coefficient is 1.35593 for all the time steps and the value of the minimum wall shear is 0 for all the time steps. In boundary-layer flows, the skin friction coefficient is an essential dimensionless parameter. It is the ratio of the local shear stress to the typical dynamics pressure, and it relates to the local value.

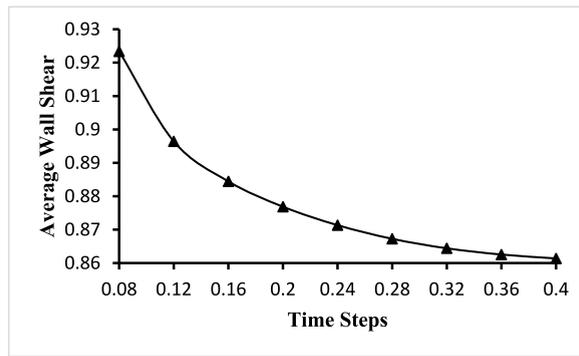

**Fig. 11.** Average wall shear variations for different time steps.

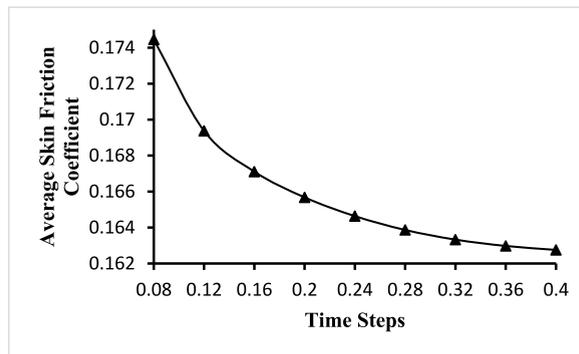

**Fig. 12.** Average skin friction coefficient variations for different time steps.

### 5.2. Hemodynamic behavior along the artery using cross-sectional planes

To study the hemodynamical behavior of the blood flow through the artery and its effects on the artery wall along the artery, we draw 30 cross-sectional planes along the artery domain. The following Fig. 13 gives the side view of the plane position. As explained in Fig. 13, we will take plane 16 as the center of the coordinate system and from here in the directions of the main carotid artery, plane 1 ($-31$ $mm$) is selected near the inlet. Again, plane 30 (18 $mm$) is selected near outlet 1 and outlet 2. The cross-sectional planes are selected to study the hemodynamic parameters including wall deformation, strain rate, flow velocity, flow pressure, wall shear, and skin friction coefficient on those planes to understand the overall behavior of those mentioned parameters along the artery.

For the analysis of the hemodynamical parameters present in the following figures (Figs. 14–19), the solid black line represents the

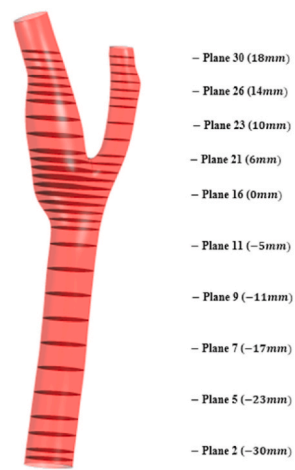

**Fig. 13.** Cross-sectional planes along the carotid artery for detailed analysis: plane 1 ($-$ 31 mm), plane 16 (0 mm), plane 30 (18 mm).

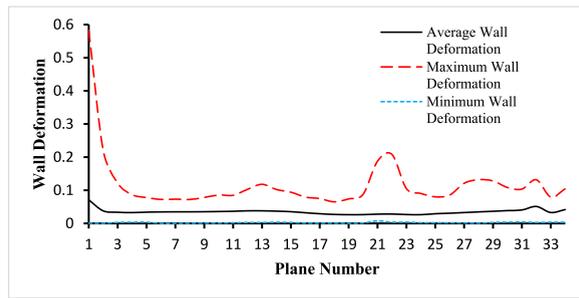

**Fig. 14.** The variations of average, maximum, and minimum wall deformation for the blood flow along the length of the carotid artery at different cross-sectional planes for t = 0.4s.

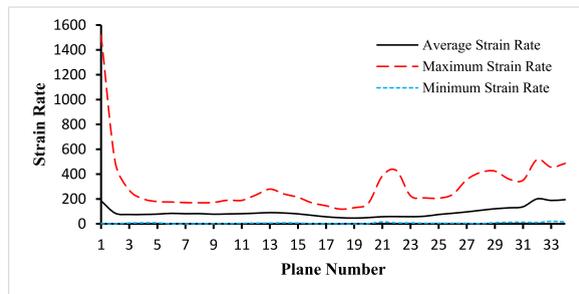

**Fig. 15.** The variations of average, maximum, and minimum strain rate for the blood flow along the length of the carotid artery at different cross-sectional planes for t = 0.4s.

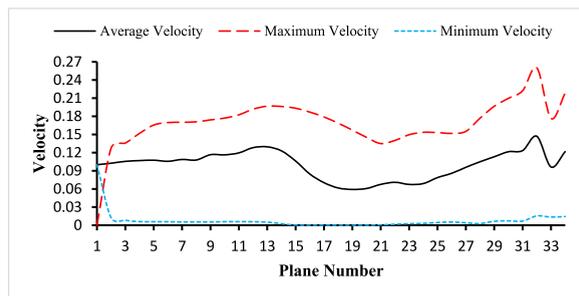

**Fig. 16.** The variations of average, maximum, and minimum velocity for the blood flow along the length of the carotid artery at different cross-sectional planes for t = 0.4s.

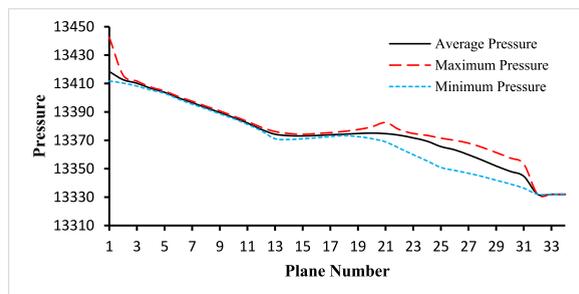

**Fig. 17.** The variations of average, maximum, and minimum pressure for the blood flow along the length of the carotid artery at different cross-sectional planes for t = 0.4s.

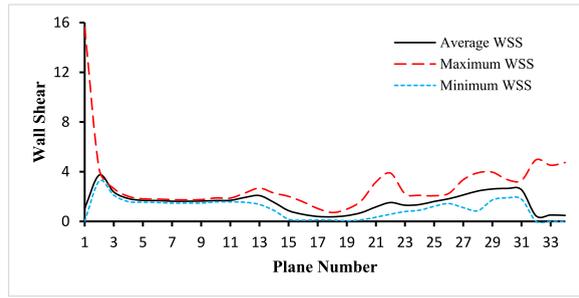

**Fig. 18.** The variations of average, maximum, and minimum wall shear stress for the blood flow along the length of the carotid artery at different cross-sectional planes for t = 0.4s.

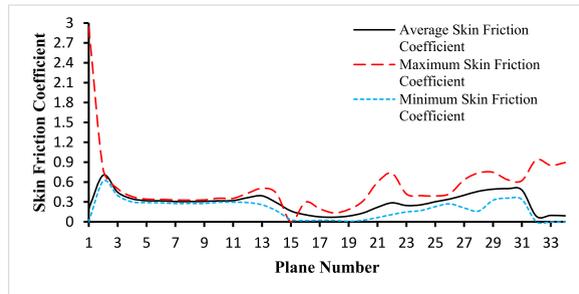

**Fig. 19.** The variations of average, maximum, and minimum skin friction coefficient for the blood flow along the length of the carotid artery at different cross-sectional planes for t = 0.4s.

average parameter value, the long dash red line represents the maximum parameter value, and the square dot light blue represents the minimum parameter value. The anemic blood type has been chosen to complete this part of the study, variations of hemodynamic parameters for different blood types are very small and those changes have been framed in the next part of the analysis. The average values have been considered for the planes in the bifurcated region (internal and external carotid artery) including plane 21 to plane 30.

Fig. 14 demonstrates the variations of the average, maximum, and minimum wall deformation for the blood flow along the length of the carotid artery at different cross-sectional planes for $t = 0.4s$. The highest value of the average wall deformation is 0.040520028 observed at plane 30 and the lowest value of the average wall deformation is 0.026286653 observed at plane 18. The highest value of the maximum wall deformation is 0.220288982 observed at plane 1 and the lowest value of the maximum wall deformation is 0.06509371 observed at plane 17. And the highest value of the minimum wall deformation is 0.007515643 observed at plane 20 and the lowest value of the minimum wall deformation is $6.4 \times 10^{-5}$ observed at plane 5. The average inlet wall deformation is 0.070953931, the average outlet 1 wall deformation is 0.051086263, and the average outlet 2 wall deformation is 0.032870549.

The variations of the average, maximum, and minimum strain rate for the blood flow along the length of the carotid artery at different cross-sectional planes for $t = 0.4s$ are given the Fig. 15. The highest value of the average strain rate is $137.324\ s^{-1}$ observed at plane 30 and the lowest value of the average strain rate is $46.1477\ s^{-1}$ observed at plane 18. The highest value of the maximum strain rate is $498.202\ s^{-1}$ observed at plane 1 and the lowest value of the maximum strain rate is $118.372\ s^{-1}$ observed at plane 17. And the highest value of the minimum strain rate is $15.4205\ s^{-1}$ observed at plane 20 and the lowest value of the minimum strain rate is $0.155507\ s^{-1}$ observed at plane 5. The average inlet strain rate is $184.798\ s^{-1}$, the average outlet 1 strain rate is $200.533\ s^{-1}$, and the average outlet 2 strain rate is $188.159\ s^{-1}$.

Fig. 16 demonstrates the variations of the average, maximum, and minimum velocity for the blood flow along the length of the carotid artery at different cross-sectional planes for $t = 0.4s$. The highest value of the average velocity is $0.129607\ ms^{-1}$ observed at plane 12 and the lowest value of the average velocity is $0.0591702\ ms^{-1}$ observed at plane 18. The highest value of the maximum velocity is $0.222315\ ms^{-1}$ observed at plane 30 and the lowest value of the maximum velocity is $0.127749\ ms^{-1}$ observed at plane 1. And the highest value of the minimum velocity is $0.0122411\ ms^{-1}$ observed at plane 1 and the lowest value of the minimum velocity is $7.11 \times 10^{-6}\ ms^{-1}$ observed at plane 18. The average inlet velocity is $0.1\ ms^{-1}$, the average outlet 1 velocity is $0.147019\ ms^{-1}$, and the average outlet 2 velocity is $0.0963011\ ms^{-1}$.

The variations of the average, maximum, and minimum pressure for the blood flow along the length of the carotid artery at different cross-sectional planes for $t = 0.4s$ are given the Fig. 17. The highest value of the average pressure is $13412.7\ Pa$ observed at plane 1 and the lowest value of the average pressure is $13344.7\ Pa$ observed at plane 30. The highest value of the maximum pressure is $13415.9\ Pa$ observed at plane 1 and the lowest value of the maximum pressure is $13353.3\ Pa$ observed at plane 30. And the highest value of the minimum pressure is 13410.3 $Pa$ observed at plane 1 and the lowest value of the minimum pressure is 13336.3 $Pa$ observed at plane 30. The average inlet pressure is 13418.4 $Pa$, the average outlet 1 pressure is 13332 $Pa$, and the average outlet 2 pressure is 13332 $Pa$.

Fig. 18 demonstrates the variations of the average, maximum, and minimum wall shear for the blood flow along the length of the carotid artery at different cross-sectional planes for $t = 0.4s$. The highest value of the average wall shear is 3.72722 $Pa$ observed in plane 1 and the lowest value of the average wall shear is 0.367042 $Pa$ observed in plane 17. The highest value of the maximum wall shear is 4.18777 $Pa$ observed in plane 1 and the lowest value of the maximum wall shear is 0.71421 $Pa$ observed in plane 17. And the highest value of the minimum wall shear is 3.24952 $Pa$ observed at plane 1 and the lowest value of the minimum wall shear is 0.0481984 $Pa$ observed at plane 18. The average inlet wall shear is 1.04676 $Pa$, the average outlet 1 wall shear is 0.445542 $Pa$, and the average outlet 2 wall shear is 0.513835 $Pa$.

The variations of the average, maximum, and minimum skin friction coefficient for the blood flow along the length of the carotid artery at different cross-sectional planes for $t = 0.4s$ are given the Fig. 19. The highest value of the average skin friction coefficient is 0.703249 observed at plane 1 and the lowest value of the average skin friction coefficient is 0.0692532 observed at plane 17. The highest value of the maximum skin friction coefficient is 0.790145 observed at plane 1 and the lowest value of the maximum skin friction coefficient is 0.134757 observed at plane 17. And the highest value of the minimum skin friction coefficient is 0.613116 observed at plane 1 and the lowest value of the minimum skin friction coefficient is 0.00909404 observed at plane 18. The average inlet skin friction coefficient is 0.197502, the average outlet 1 skin friction coefficient is 0.0840645, and the average outlet 2 skin friction coefficient is 0.09695.

*5.3. Effects of different blood types on the hemodynamic behavior*

For the study of the different blood types as mentioned in Table 1, this section analyzes the 3$D$ view of the variations of the velocity streamlines, velocity vectors, pressure contours, wall shear contours, and strain rate contours with detailed explanations. The variational effects have been calculated for the anemic, diabetic, and two healthy cases of blood types. To understand the specific variations of the hemodynamic parameters at the different locations through the artery, the following Table 2 considers the seven planes through the domain. Variational effects of the average value, maximum value, and minimum value have been compared for each hemodynamic parameter mentioned in the following figures.

Fig. 20 shows the effects of the anemic, diabetic, healthy (case 1), and healthy (case 2) blood flow on the velocity streamlines through the artery at time step $t = 0.4s$. The four sub-parts of the figure show the variations for the four different blood cases as mentioned. To understand the specific velocity values at different positions of the artery, we consider Fig. 21 to compare the average velocity, maximum velocity, and minimum velocity variations of the seven planes as mentioned in Table 2.

In this part, the solid line, long dash line, dash-dot line, and the dashed line represent the properties of the anemic, diabetic, healthy (case 1), and healthy (case 2) respectively. The velocity values are given by the four black lines and the red and blue lines represent the trendlines for corresponding blood types. Fig. 21(a) shows the variations of the average velocities calculated for the different blood cases at the seven planes as mentioned in Table 2. From the linear trendline values in Fig. 21(a), we can see that the average velocity values for anemic and diabetic blood are lower than the values for the healthy cases. The highest value of the average blood velocity is 0.1261 $ms^{-1}$ obtained for the healthy (case 1) blood types at the outlet and the lowest value of the average blood velocity is 0.0598 $ms^{-1}$ obtained for the healthy (case 2) blood types at the plane 19. The lowest average velocity at the outlet (average of outlet 1 and outlet 2) is 0.12165 $ms^{-1}$ obtained for diabetic blood types. Fig. 21(b) shows the variations of the maximum velocities calculated for the different blood cases at the seven planes as mentioned in Table 2. From the linear trendline values in Fig. 21(b), we can see that the maximum velocity values for the anemic and diabetic blood are lower than the values for the healthy (case 1) but higher than the healthy (case 2) blood types. The highest value of the maximum blood velocity is 0.2184 $ms^{-1}$ obtained for the diabetic blood type at the outlet and the lowest value of the maximum blood velocity is 0.140189 $ms^{-1}$ obtained for the anemic blood types at plane 21. The lowest maximum velocity at the outlet (average of outlet 1 and outlet 2) is 0.20715 $ms^{-1}$ obtained for healthy (case 2) blood types. Fig. 21(c) shows the variations of the minimum velocities calculated for the different blood cases at the seven planes as mentioned in Table 2. From the linear trendline values in Fig. 21(c), we can see that the minimum velocity values for anemic and diabetic blood are lower than the values for the healthy blood type cases. The highest value of the minimum blood velocity is 0.017105 $ms^{-1}$ obtained for the healthy (case 2) blood type at the outlet and the lowest value of the maximum blood velocity is $5.32 \times 10^{-5}$ $ms^{-1}$ obtained also for the healthy (case 2) blood type at plane 21. The lowest minimum velocity at the outlet (average of outlet 1 and outlet 2) is 0.01449 $ms^{-1}$ obtained for the diabetic blood types.

The rate of change of an object's location is represented by a velocity vector. The magnitude of a velocity vector indicates an object's speed, whereas the vector direction indicates the object's direction. The effects of the anemic, diabetic, healthy (case 1), and healthy (case 2) blood flow on the velocity vectors through the artery at time step $t = 0.4s$ is shown in Fig. 22.

Fig. 23 shows the effects of the anemic, diabetic, healthy (case 1), and healthy (case 2) blood flow on the pressure contours through the artery at time step $t = 0.4s$. The four sub-parts of the figure show the variations for the four different blood cases as mentioned. To understand the specific pressure values at different positions of the artery, we consider Fig. 24 to compare the average pressure, maximum pressure, and minimum pressure variations of the seven planes as mentioned in Table 2.

**Table 2**
Marking of the seven sample planes for detailed analysis.

| Mark | 1 | 2 | 3 | 4 | 5 | 6 | 7 |
|---|---|---|---|---|---|---|---|
| **Plane Number** | Plane 4 | Plane 15 | Plane 19 | Plane 21 | Plane 23 | Plane 28 | Outlet |

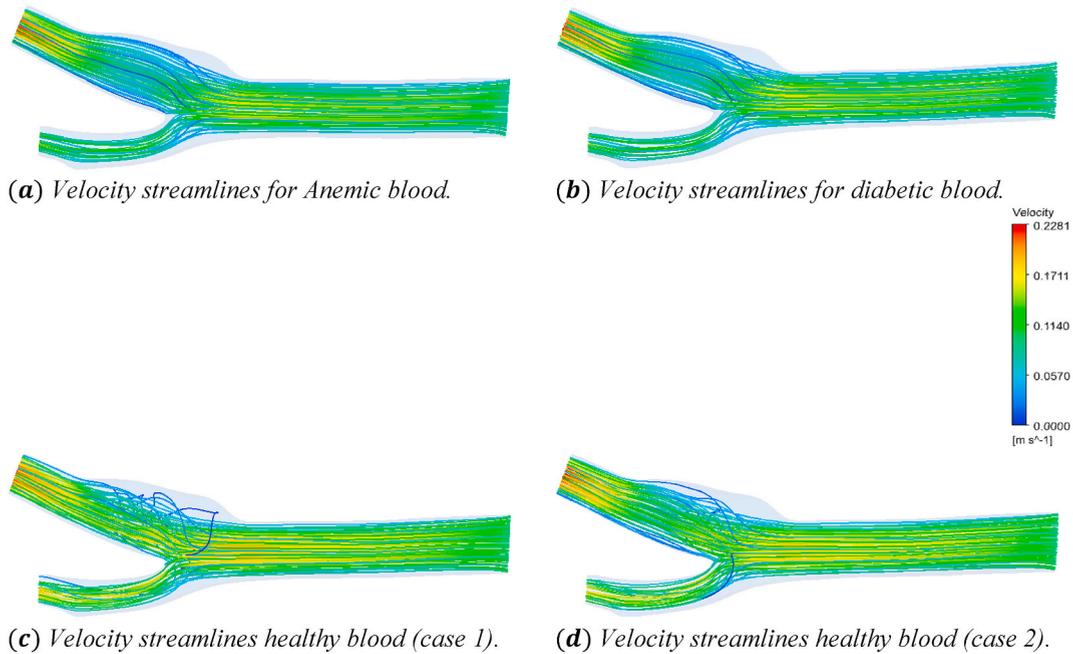

**Fig. 20.** The effect of Anemic (a), diabetic (b), and healthy (case 1 (c) and case 2 (d)) blood flow on the velocity streamlines in the carotid artery at t = 0.4s.

In this part, the solid line, long dash line, dash-dot line, and the dashed line represent the properties of the anemic, diabetic, healthy (case 1), and healthy (case 2) respectively. The pressure values are given by the four black lines and the red and blue lines represent the trendlines for corresponding blood types. Fig. 24(a) shows the variations of the average pressure values calculated for the different blood cases at the seven planes as mentioned in Table 2. From the linear trendline values in Fig. 24 (a), we can see that the average pressure values for anemic and diabetic blood are higher than the values for the healthy cases. The highest value of the average blood pressure is 13403.8 *Pa* obtained for the anemic blood types at plane 4 and the lowest value of the average blood pressure is 13330 *Pa* obtained for both healthy case blood types at the outlet. We can observe that the average blood pressure for the animatic and diabetic blood types in the outlet is higher than the blood pressures measured for the healthy cases. Fig. 24(b) shows the variations of the maximum pressure values calculated for different blood cases at the seven planes as mentioned in Table 2. From the linear trendline values in Fig. 24(b), we can see that the maximum pressure values for anemic and diabetic blood are higher than the values for the healthy cases. The highest value of the maximum blood pressure is 13404.7 *Pa* obtained for the anemic blood types at plane 4 and the lowest value of the maximum blood pressure is 13330 *Pa* obtained for both healthy case blood types at the outlet. We can observe that the maximum blood pressure for the animatic and diabetic blood types in the outlet is higher than the blood pressures measured for the healthy cases. Fig. 24(c) shows the variations of the minimum pressure values calculated for different blood cases at the seven planes as mentioned in Table 2. From the linear trendline values in Fig. 24(c), we can see that the minimum pressure values for anemic and diabetic blood are higher than the values for the healthy cases. The highest value of the minimum blood pressure is 13403.1 *Pa* obtained for the anemic blood types at plane 4 and the lowest value of the minimum blood pressure is 13330 *Pa* obtained for both healthy case blood types at the outlet. We can observe that the minimum blood pressure for the animatic and diabetic blood types in the outlet is higher than the blood pressures measured for the healthy cases.

Fig. 25 shows the effects of the anemic, diabetic, healthy (case 1), and healthy (case 2) blood flow on the wall shear contours through the artery at time step $t = 0.4s$. The four sub-parts of the figure show the variations for the four different blood cases as mentioned. To understand the specific wall shear values at different positions of the artery, we consider Fig. 26 to compare the average wall shear, maximum wall shear, and minimum wall shear variations of the seven planes as mentioned in Table 2.

In this part, the solid line, long dash line, dash-dot line, and the dashed line represent the properties of the anemic, diabetic, healthy (case 1), and healthy (case 2) respectively. The wall shear values are given by the four black lines and the red and blue lines represent the trendlines for corresponding blood types. Fig. 26(a) shows the variations of the average wall shear values calculated for the different blood cases at the seven planes as mentioned in Table 2. From the linear trendline values in Fig. 26(a), we can see that the average wall shear values for anemic and diabetic blood are higher than the values for the healthy cases. The highest value of the average wall shear is 2.608 *Pa* obtained for the diabetic blood types at plane 28 and the lowest value of the average wall shear is 0.1714 *Pa* obtained for the healthy (case 1) blood types at plane 15. The highest value of the average wall shear at the outlet is

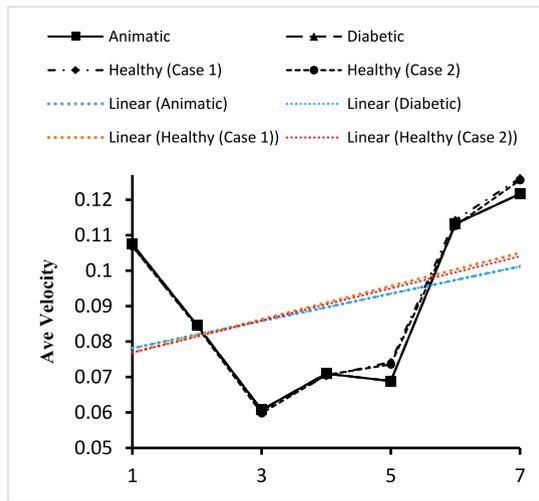

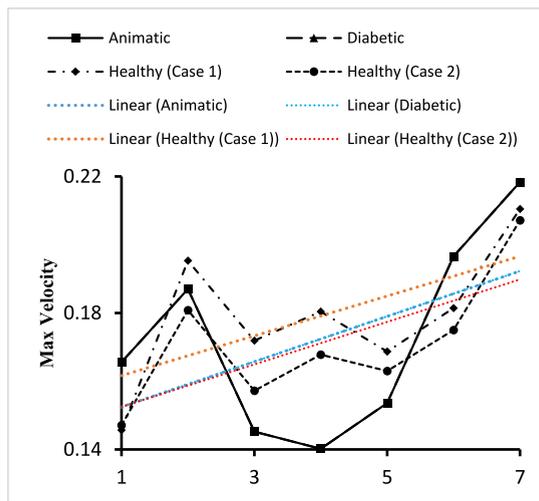

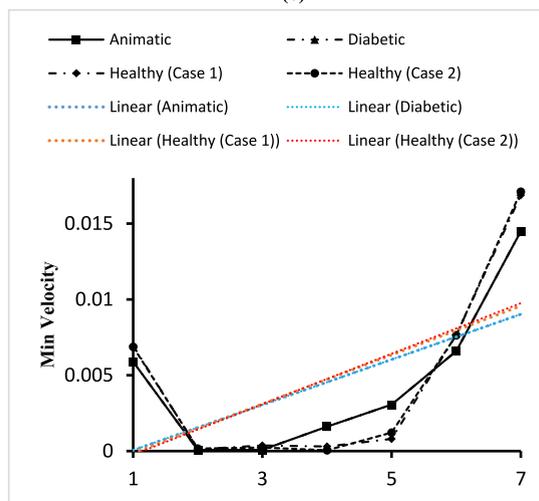

**Fig. 21.** (a). Variations of the average velocity for the four blood cases calculated at planes 4, 15, 19, 21, 23, and 28.
(b). Variations of the maximum velocity for the four blood cases calculated at planes 4, 15, 19, 21, 23, and 28.
(c). Variations of the minimum velocity for the four blood cases calculated at planes 4, 15, 19, 21, 23, and 28.

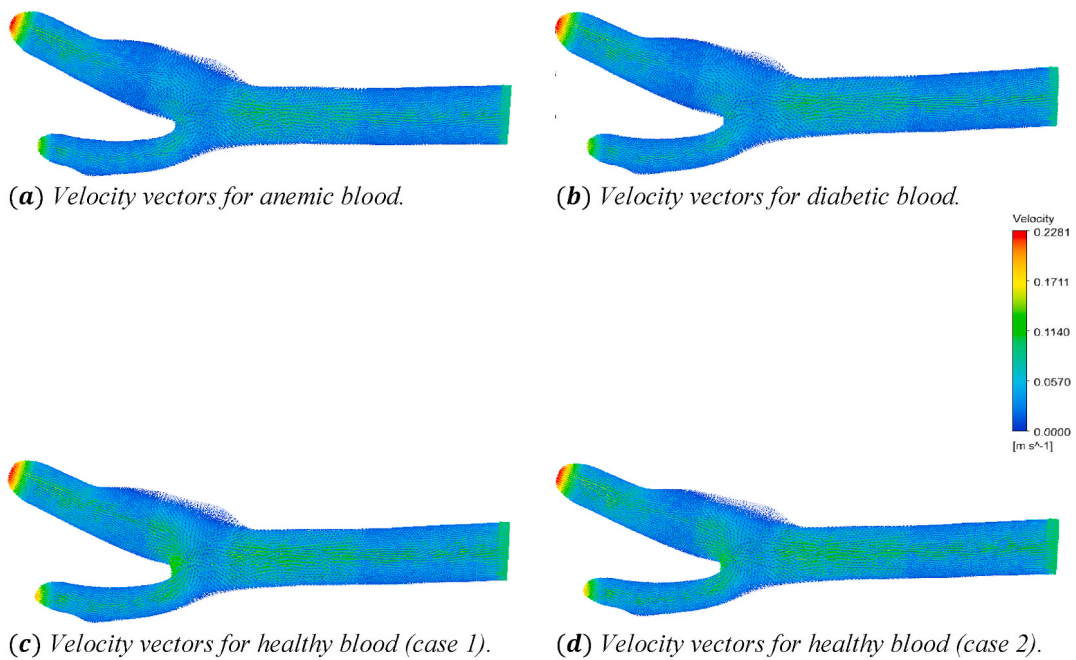

**Fig. 22.** The effect of Anemic (a), diabetic (b), and healthy (case 1 (c) and case 2 (d)) blood flow on the velocity vectors in the carotid artery at t = 0.4s.

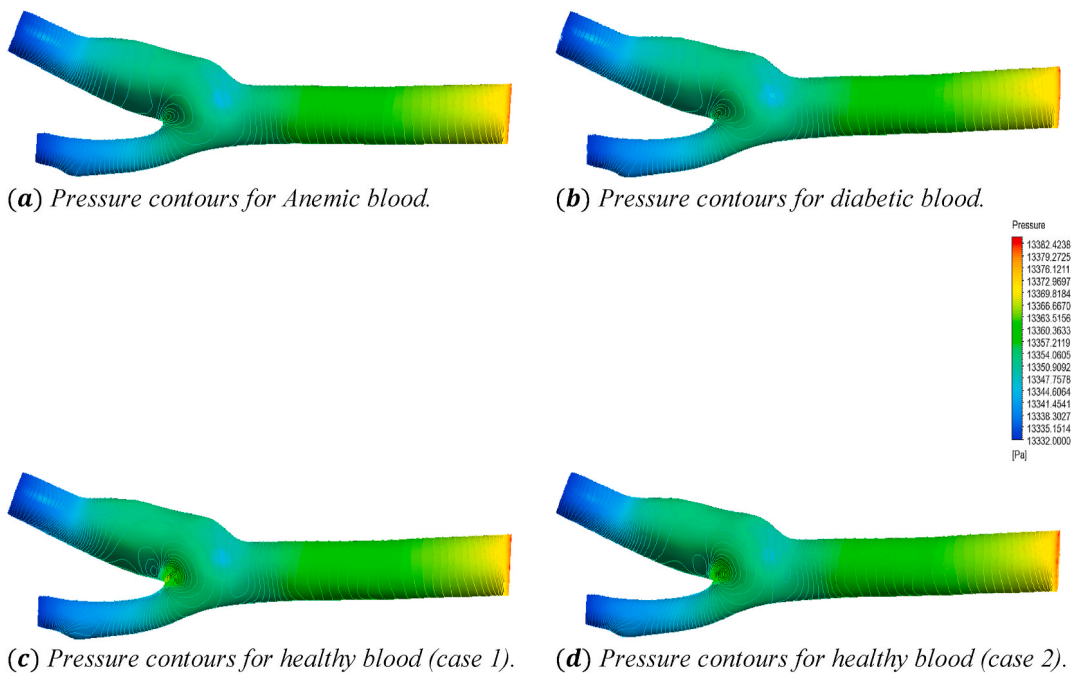

**Fig. 23.** The effect of Anemic (a), diabetic (b), and healthy (case 1 (c) and case 2 (d)) blood flow on the pressure contours in the carotid artery walls (3D) at t = 0.4s.

0.2551 *Pa* for the healthy (case 2) blood type and the lowest value of the average wall shear at the outlet is 0.47965 *Pa* for the diabetic blood type. Fig. 26(b) shows the variations of the maximum wall shear values calculated for different blood cases at the seven planes as mentioned in Table 2. From the linear trendline values in Fig. 26(b), we can see that the maximum wall shear values for anemic and diabetic blood are higher than the values for the healthy cases. The highest value of the maximum wall shear is 4.738715 *Pa* obtained for the anemic blood types at the outlet and the lowest value of the maximum wall shear is 0.6511 *Pa* obtained for the healthy (case 1) blood types at plane 15. The highest value of the maximum wall shear at the outlet is 4.738715 $Pa$ for the anemic blood type and the lowest value of the maximum wall shear at the outlet is 2.4705 $Pa$ for the healthy (case 1) blood type. Fig. 26(c) shows the variations of the minimum wall shear values calculated for the different blood cases at the seven planes as mentioned in Table 2. From the linear trendline values in Fig. 26(c), we can see that the minimum wall shear values for anemic and diabetic blood are higher than the values for the healthy cases. The highest value of the minimum wall shear is 1.72449 $Pa$ obtained for the anemic blood types at plane 28 and the lowest value of the minimum wall shear is 0 $Pa$ obtained for all the blood types at the outlet.

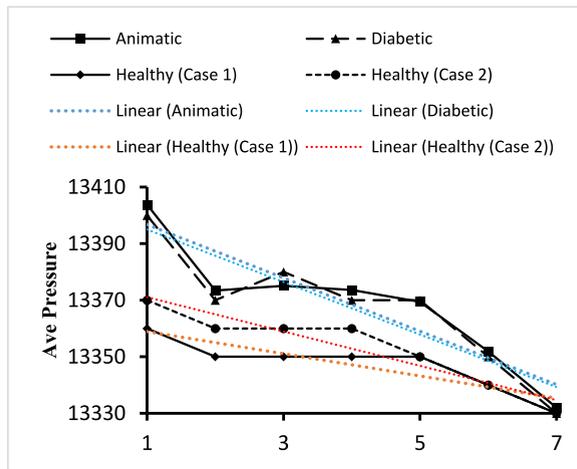

(a)

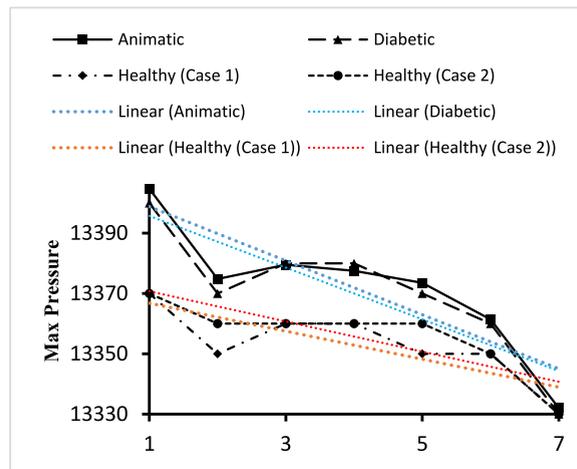

(b)

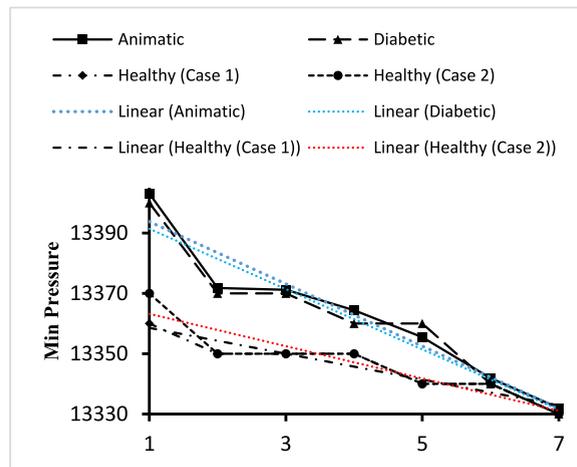

(c)

**Fig. 24.** (a). Variations of the average pressure for the four blood cases calculated at the planes 4, 15, 19, 21, 23, and 28.
(b). Variations of the maximum pressure for the four blood cases calculated at planes 4, 15, 19, 21, 23, and 28.
(c). Variations of the minimum pressure for the four blood cases calculated at planes 4, 15, 19, 21, 23, and 28.

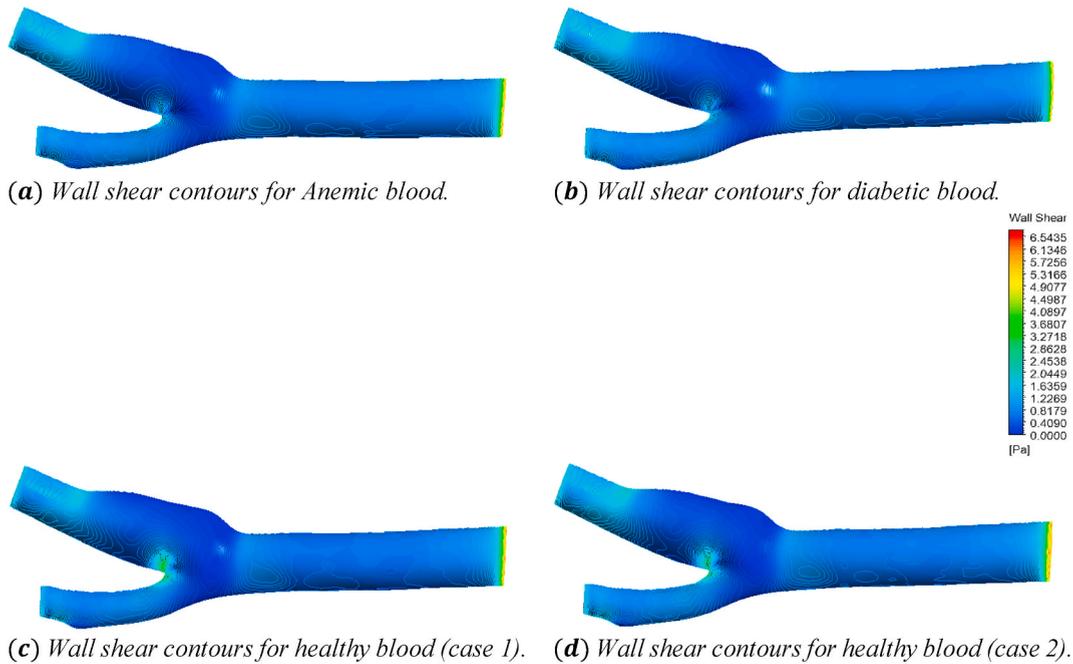

**Fig. 25.** The effect of Anemic (a), diabetic (b), and healthy (case 1 (c) and case 2 (d)) blood flow on the wall shear contours in the carotid artery walls (3D) at t = 0.4s.

Fig. 27 shows the effects of the anemic, diabetic, healthy (case 1), and healthy (case 2) blood flow on the strain rate contours through the artery at time step $t = 0.4s$. The four sub-parts of the figure show the variations for the four different blood cases as mentioned. To understand the specific strain rate values at different positions of the artery, we consider Fig. 27 to compare the average strain rate, maximum strain rate, and minimum strain rate variations of the seven planes as mentioned in Table 2.

In the following part, the solid line, long dash line, dash-dot line, and the dashed line represent the properties of the anemic, diabetic, healthy (case 1), and healthy (case 2) respectively. The strain rate values are given by the four black lines and the red and blue lines represent the trendlines for corresponding blood types.new. Fig. 28(a) shows the variations of the average strain rate values calculated for different blood cases at the seven planes as mentioned in Table 2. From the linear trendline values in Fig. 28(a), we can see that the average strain rate values for anemic and diabetic blood are lower than the values for the healthy cases. The highest value of the average strain rate is 204.15 $s^{-1}$ obtained for the healthy (case 1) blood types at the outlet and the lowest value of the average strain rate is 49.99 $s^{-1}$ obtained for the diabetic blood types at plane 19. The highest value of the average strain rate at the outlet is 204.15 $s^{-1}$ for the healthy (case 1) blood type and the lowest value of the average strain rate at the outlet is 194.346 $s^{-1}$ for the anemic blood type. Fig. 28(b) shows the variations of the maximum strain rate values calculated for the different blood cases at the seven planes as mentioned in Table 2. From the linear trendline values in Fig. 28(b), we can see that the maximum strain rate values for the anemic and diabetic blood are lower than the values for the healthy (case 1) blood type but higher than the healthy (case 2) blood type. The highest value of the maximum strain rate is 912 $s^{-1}$ obtained for the healthy (case 1) blood types at plane 21 and the lowest value of the average strain rate is 165.784 $s^{-1}$ obtained for the anemic blood types at plane 19. The highest value of the maximum strain rate at the outlet is 646.45 $s^{-1}$ for the healthy (case 1) blood type and the lowest value of the maximum strain rate at the outlet is 486.3885 $s^{-1}$ for the anemic blood type. Fig. 28(c) shows the variations of the minimum strain rate values calculated for different blood cases at the seven planes as mentioned in Table 2. From the linear trendline values in Fig. 28(c), we can see that the average strain rate values for anemic and diabetic blood are lower than the values for the healthy cases.

The highest value of the minimum strain rate is 17.055 $s^{-1}$ obtained for the healthy (case 1) blood types at the outlet and the lowest value of the average strain rate is 1.195 $s^{-1}$ obtained for the healthy (case 2) blood types at plane 21. The highest value of the average strain rate at the outlet is 17.055 $s^{-1}$ for the healthy (case 1) blood type and the lowest value of the average strain rate at the outlet is 15.385 $s^{-1}$ for the diabetic blood type.

As a result, it is deduced that various blood types have a major impact on blood flow and arterial hemodynamics. The following

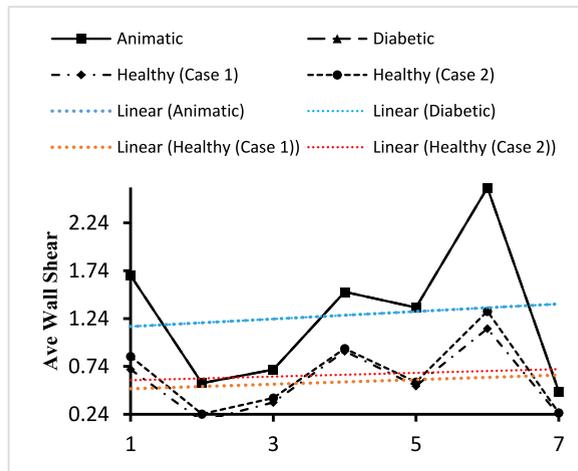

(a)

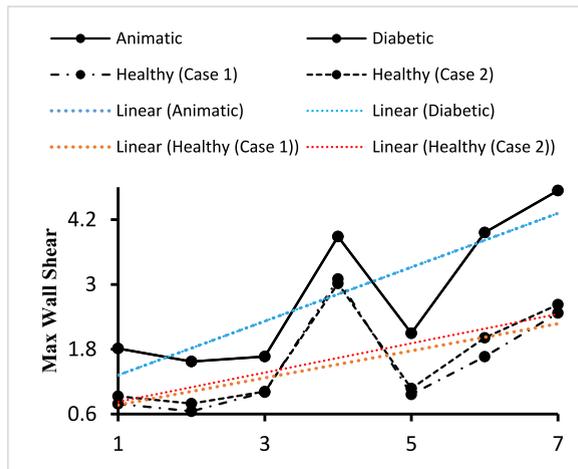

(b)

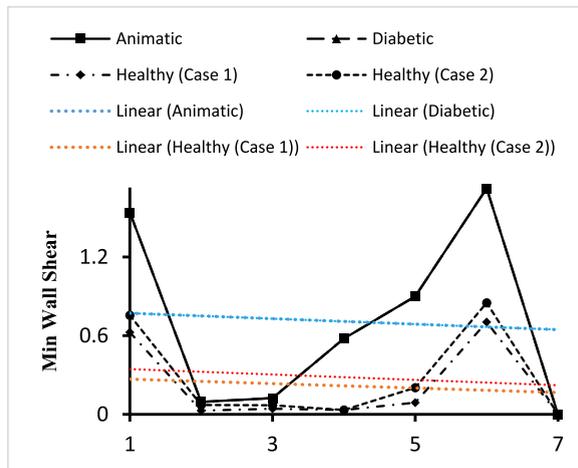

(c)

**Fig. 26.** (a). Variations of the average wall shear for the four blood cases calculated at planes 4, 15, 19, 21, 23, and 28.
(b). Variations of the maximum wall shear for the four blood cases calculated at planes 4, 15, 19, 21, 23, and 28.
(c). Variations of the minimum wall shear for the four blood cases calculated at planes 4, 15, 19, 21, 23, and 28.

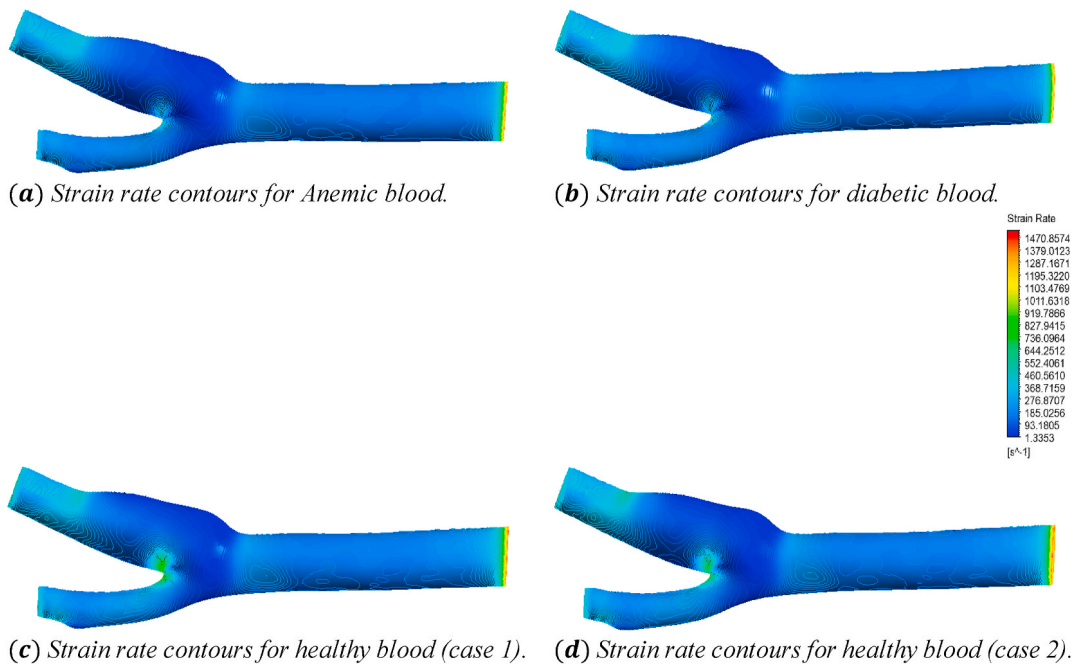

**Fig. 27.** The effect of Anemic (a), diabetic (b), and healthy (case 1 (c) and case 2 (d)) blood flow on the strain rate contours in the carotid artery walls (3D) at t = 0.4s.

Table 3 is represented by categorizing the changing effects into high, average, and low on a comparative view after monitoring the influence of the four distinct blood types on the velocity, pressure, wall shear, and strain rate along the carotid artery.

## 6. Conclusion

This paper focuses on the hemodynamic behavior study in the carotid artery for anemic, diabetic, and two healthy blood instances. The external and internal carotid arteries are bifurcated in the common carotid artery, and the viscosity of various blood instances has been determined using the Carreau model. Several cross-sectional planes along the artery were used to study the impact of time steps and four blood types. The following are the specific findings of this investigation:

- The average wall deformation, average strain rate, and blood velocity all rise when the time steps are increased, while arterial pressure and wall shear drop.
- The average wall deformation, arterial pressure, wall shear, and skin friction coefficient decrease as the blood velocity increases through the artery from the common carotid artery to the outlet.
- The blood velocity of anemic and diabetic patients is lower than that of healthy patients in the maximum, average, and minimum velocity situations.
- Anemic and diabetic blood cases have greater arterial pressure and wall shear than healthy blood instances.
- Although healthy case 2 shows mixed behavior, the strain rate for anemic and diabetic blood is lower than for healthy case 1.

For future development on this project, the following recommendation might be made:

- The impacts on hemodynamic behaviors may be calculated using the stenosed artery.
- Improvements to the boundary conditions of both the fluid and solid domains are required. The vessel wall's permanent inlet and outlet rings contribute artificial wave reflections to the construction, which should be considered.
- It will be possible to conclude the effect of cardiovascular artery associated with stroke mechanism.
- The correlation between artery wall shear stress and stroke occurrence will be analyzed.
- The study will be extended to analyze atherosclerosis using hemodynamical parameters such as time dependent wall shear stress and oscillatory shear stress index.

**Data availability statement**

The data that support the findings of this study are available online at https://github.com/Hashnayne-Ahmed/blood_flow.

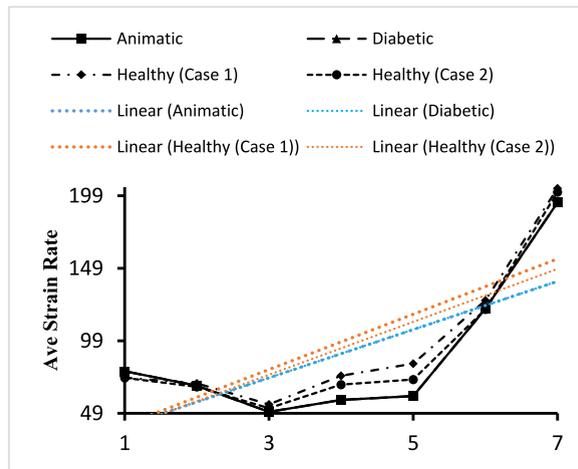

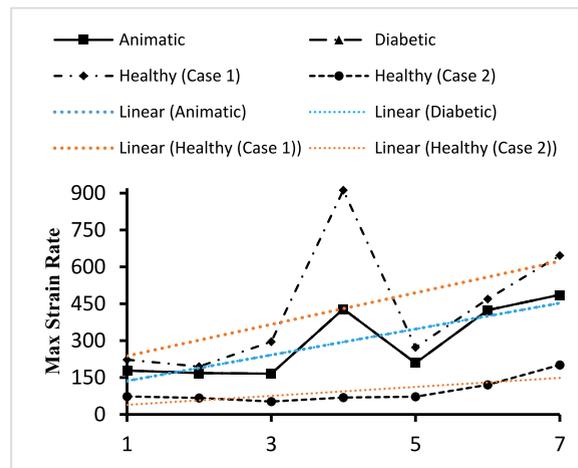

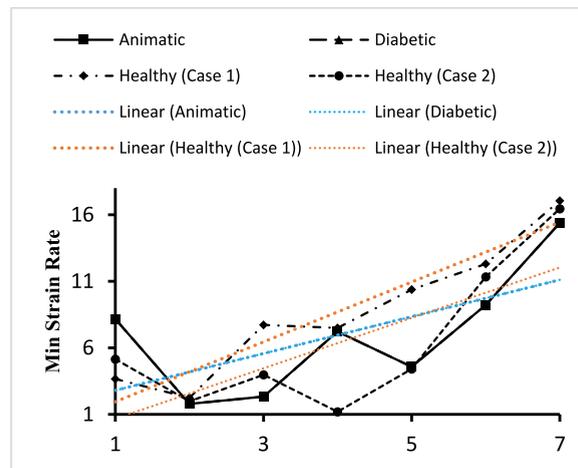

**Fig. 28.** (a). Variations of the average strain rate for the four blood cases calculated at the planes 4, 15, 19, 21, 23, and 28.
(b). Variations of the maximum strain rate for the four blood cases calculated at planes 4, 15, 19, 21, 23, and 28.
(c). Variations of the minimum strain rate for the four blood cases calculated at planes 4, 15, 19, 21, 23, and 28.

**Table 3**

Comparative hemodynamic behavior analysis for different blood types in the carotid artery.

| Hemodynamic Parameters | | Anemic | Diabetic | Healthy (Case 1) | Healthy (Case 2) |
|---|---|---|---|---|---|
| **Blood Velocity** | Maximum | average | average | high | low |
| | Average | low | low | high | high |
| | Minimum | low | low | high | high |
| **Arterial Pressure** | Maximum | high | high | low | low |
| | Average | high | high | low | average |
| | Minimum | high | high | low | low |
| **Wall Shear** | Maximum | high | high | low | low |
| | Average | high | high | low | low |
| | Minimum | high | high | low | low |
| **Strain Rate** | Maximum | average | average | high | low |
| | Average | low | low | high | high |
| | Minimum | average | average | high | low |

**Author contributions**

**Hashnayne Ahmed:** Writing – review & editing, Writing – original draft, Visualization, Validation, Software, Resources, Methodology, Investigation, Funding acquisition, Formal analysis, Data curation, Conceptualization. **Chinmayee Podder:** Writing – review & editing, Validation, Supervision, Project administration, Methodology, Funding acquisition, Conceptualization.

**Declaration of competing interest**

The authors declare that they have no known competing financial interests or personal relationships that could have appeared to influence the work reported in this paper.

**Acknowledgments**

This work is partially supported by the National Science & Technology (NST) fellowship program of the Ministry of Science and Technology, People's Republic of Bangladesh (Number – 2021/39.00.0000.012.05.20–05, Registration Number – 949).

**References**

[1] Brenda R. Kwak, et al., Biomechanical factors in atherosclerosis: mechanisms and clinical implications, Eur. Heart J. 35 (43) (2014) 3013–3020.
[2] Bernard Fox, et al., Distribution of fatty and fibrous plaques in young human coronary arteries, Atherosclerosis 41.2–3 (1982) 337–347.
[3] Michael Eliasziw, et al., Early risk of stroke after a transient ischemic attack in patients with internal carotid artery disease, CMAJ (Can. Med. Assoc. J.) 170 (7) (2004) 1105–1109.
[4] Hugh Markus, Marisa Cullinane, Severely impaired cerebrovascular reactivity predicts stroke and TIA risk in patients with carotid artery stenosis and occlusion, Brain 124 (3) (2001) 457–467.
[5] P.G. Bakker, Bifurcations in Flow Patterns, Nonlinear topics in the mathematical sciences, 1991.
[6] Awad Javaid, Mostafa Alfishawy, "Internal Carotid Artery Stenosis Presenting with Limb Shaking TIA." *Case Reports in Neurological Medicine* 2016, 2016.
[7] Maria Jeziorska, David E. Woolley, Local neovascularization and cellular composition within vulnerable regions of atherosclerotic plaques of human carotid arteries, J. Pathol. 188 (2) (1999) 189–196.
[8] Norbert Nighoghossian, Laurent Derex, Douek Philippe, The vulnerable carotid artery plaque: current imaging methods and new perspectives, Stroke 36 (12) (2005) 2764–2772.
[9] Ryo Torii, et al., Fluid-structure interaction modeling of aneurysmal conditions with high and normal blood pressures, Comput. Mech. 38 (4) (2006) 482–490.
[10] Y.C. Fung, Skalak Richard, Biomechanics: mechanical properties of living tissues (1981) 231–298.
[11] J.D. Humphrey, Cardiovascular Solid Mechanics: Cells, Tissues, and Organs, 35, Springer-Verlag., New York, 2002.
[12] Ryo Torii, et al., Influence of wall elasticity in patient-specific hemodynamic simulations, Comput. Fluid 36 (1) (2007) 160–168.
[13] Seungik Baek, et al., Theory of small on large: potential utility in computations of fluid-solid interactions in arteries, Comput. Methods Appl. Mech. Eng. 196 (31–32) (2007) 3070–3078.
[14] J.D. Humphrey, Cardiovascular Solid Mechanics: Cells, Tissues, and Organs, 35, Springer-Verlag., New York, 2002.
[15] T. Christian Gasser, Gerhard A. Holzapfel, Finite element modeling of balloon angioplasty by considering overstretch of remnant non-diseased tissues in lesions, Comput. Mech. 40 (1) (2007) 47–60.
[16] G. Mengaldo, et al., A Comparative Study of Different Nonlinear Hyperelastic Isotropic Arterial Wall Models in Patient-specific Vascular Flow Simulations in the Aortic Arch, 2012.
[17] Yuri Bazilevs, Kenji Takizawa, Tayfun E. Tezduyar, Computational Fluid-Structure Interaction: Methods and Applications, John Wiley & Sons, 2013.
[18] Wook Yang Joung, et al., Wall shear stress in hypertensive patients is associated with carotid vascular deformation assessed by speckle tracking strain imaging, Clin. Hyperten. 20 (1) (2014) 1–6.
[19] Samaee Milad, Tafazzoli-Shadpour Mohammad, Alavi Hamed, Coupling of shears–circumferential stress pulses investigation through stress phase angle in FSI models of the stenotic artery using experimental data, Med. Biol. Eng. Comput. 55 (8) (2017) 1147–1162.
[20] Mandy L. Corrigan, Handbook of Clinical Nutrition and Stroke, Springer, New York, 2013.
[21] Mozaffarian Dariush, et al., Heart disease and stroke statistics—2016 update: a report from the American Heart Association, Circulat 133 (4) (2016) e38–e360.
[22] J.C. Barbenel, "The Measurement of Red Blood Cell deformability." Clinical Aspects of Blood Viscosity and Cell Deformability, Springer, London, 1981, pp. 37–45.
[23] Concetta Irace, et al., Blood viscosity in subjects with normoglycemia and prediabetes, Diabet. Care 37 (2) (2014) 488–492.
[24] Pankaj Mathur, Surekha Jain, Mathematical modeling of non-Newtonian blood flow through an artery in the presence of stenosis, Adv. Appl. Math. Biosci. 4 (1) (2013) 1–12.

[25] Hieltje Goslinga, Blood Viscosity and Shock: the Role of Hemodilution, Hemoconcentration, and Defibrination, vol. 160, Springer Science & Business Media, 2012.
[26] Ioana Mozos, Mechanisms linking red blood cell disorders and cardiovascular diseases, BioMed Res. Int. (2015) 2015.
[27] Mandy L. Corrigan, Handbook of Clinical Nutrition and Stroke, Springer, New York, 2013.
[28] Siebert Mark W., Fodor Petru S., Newtonian and Non-Newtonian Blood Flow Over a Backward-Facing Step–A Case Study. Proceedings of the COMSOL Conference, Boston, 2009.
[29] Pauli Lutz, Behr Marek, On stabilized space-time FEM for anisotropic meshes: incompressible Navier–Stokes equations and applications to blood flow in medical devices, Int. J. Numer. Methods Fluid. 85 (3) (2017) 189–209.
[30] Paula J. Richards, "Susan Standing (Editor-In-Chief), Grays Anatomy: the Anatomical Base of Clinical Practice—39th Edition, Elsevier Churchill Livingstone, 2006, p. 472.
[31] Jaroslaw Krejza, et al., Carotid artery diameter in men and women and the relation to body and neck size, Stroke 37 (4) (2006) 1103–1105.
[32] Harunobu Shima, et al., Anatomy of microvascular anastomosis in the neck, Plast. Reconstr. Surg. 101 (1) (1998) 33–41.
[33] Radwa M. Attia, AA Eldosoky Mohamed, Studying of the blood flow behavior in a stenosed carotid artery for healthy, anemic and diabetic blood, International Conference on Innovative Trends in Communication and Computer Engineering (ITCE) (2020) 72–75. IEEE.
[34] cornellsimulation Cornell University YouTube channel. https://www.youtube.com/watch?v=dVjMBGLzDfM. (Accessed on 28 December 2021)..
[35] Bouteloup Hugo, Johann Guimaraes de Oliveira Marinho, Daniel M., Espino, Computational analysis to predict the effect of pre-bifurcation stenosis on the hemodynamics of the internal and external carotid arteries, J. Mech. Eng. Sci. 14(3) (2020) 7029–7039.
[36] Huy Dinh, et al., Reconstruction of Carotid Stenosis Hemodynamics Based on Guidewire Pressure Data and Computational Modeling, Medical & Biological Engineering & Computing, 2022, pp. 1–16.
[37] Ke Yang, et al., Wall shear stress measurement in carotid artery phantoms with variation in degree of stenosis using plane wave vector Doppler, Appl. Sci. 13 (1) (2023) 617.
[38] Payam Jalili, et al., Analytical and numerical investigation of thermal distribution for hybrid nanofluid through an oblique artery with mild stenosis, SN Appl. Sci. 5 (4) (2023) 95.
[39] Manli Zhou, et al., Wall shear stress and its role in atherosclerosis, Frontiers in Cardiovascular Medicine 10 (2023) 1083547.
[40] A.M. Moerman, et al., The correlation between wall shear stress and plaque composition in advanced human carotid atherosclerosis, Front. Bioeng. Biotechnol. 9 (2022) 1499.
[41] Barrett Kim, et al., Ganong's Review of Medical Physiology, 26, McGraw Hill/Medical, 2019, pp. 1252–1358.
[42] E. Hall John, Michael E. Hall, Guyton and Hall Textbook of Medical Physiology E-Book, Elsevier Health Sciences 14 (2020) 167–180.